\documentclass[letterpaper,pra,aps,twocolumn,showpacs]{revtex4}
\usepackage{graphicx,bm,color}    

\begin{document}

\title{Quantum phases of quadrupolar Fermi gases in coupled one-dimensional systems}
\author{Wen-Min Huang, M. Lahrz and L. Mathey}
\affiliation{Zentrum f\"ur Optische Quantentechnologien and Institut f\"ur Laserphysik, Universit\"at Hamburg, 22761 Hamburg, Germany\\
Hamburg Centre for Ultrafast Imaging, Luruper Chaussee 149, Hamburg 22761, Germany
}

\begin{abstract} 
Following the recent proposal to create quadrupolar gases [Bhongale et al., Phys. Rev. Lett. 110, 155301 (2013)], we investigate what quantum phases can be created in these systems in one dimension. We consider a geometry of two coupled one-dimensional systems, and derive the quantum phase diagram of ultra-cold fermionic atoms interacting via quadrupole-quadrupole interaction within a Tomonaga-Luttinger-liquid framework. We map out the phase diagram as a function of the distance between the two tubes and the angle between the direction of the tubes and the quadrupolar moments. The latter can be controlled by an external field. We show that there are two magic angles $\theta^{c}_{B,1}$ and $\theta^{c}_{B,2}$ between $0$ to $\pi/2$, where the intratube quadrupolar interactions vanish and change signs. Adopting a pseudo-spin language with regards to the two 1D systems, the system undergoes a spin-gap transition and displays a zig-zag density pattern, above $\theta^{c}_{B,2}$ and below $\theta^{c}_{B,1}$. Between the two magic angles, we show that polarized  triplet superfluidity and a planar spin-density wave order compete  with each other. The latter corresponds to a bond order solid in higher dimensions. We demonstrate that this order can be further stabilized by applying a commensurate periodic potential along the tubes. 
\end{abstract}

\date{\today}

\pacs{67.85.-d, 67.85.Lm, 71.10.Pm}

\maketitle

\section{Introduction}
Throughout the development of the field of ultra-cold atoms, new features of the atomic systems have been developed and explored~\cite{Bloch08}. Since the condensation of bosonic atoms in a single hyperfine state in a three dimensional trap, the scope of the field has expanded to spinor condensates, bosonic mixtures, fermionic atoms, Bose-Fermi mixtures and atoms in optical lattices in one to three dimensions, to name but a few. A particular novel development was the study of atoms and molecules interacting via dipolar interactions~\cite{Baranov12}. The higher-order symmetry of this interaction, and the $1/r^{3}$ behavior of the potential of the distance $r$, adds an intriguing new feature to ultra-cold atom ensembles. In particular, the stabilization of pure dipolar quantum gases in recent experiments~\cite{Ni,Koch,Deiglmayr,Ni10,Lu,Chotia,Wu,Yan} has triggered numerous theoretical studies~\cite{Wang06,Micheli,Baranov,Dulieu,Carr,Lahaye}. The anistopic interaction between fermionic dipolar molecules has been predicted to drive unconventional pairing in  two-dimensional layers~\cite{Bruun,Cooper,Levinsen}. For a nested Fermi surface, such as for dipolar atoms in optical lattices, density-wave instabilities with nonzero angular momentum can dominate~\cite{Yamaguchi,Mikelsons,Parish,Bhongale1,Bhongale2,Bruun2,Block}. In a multilayered structure, interlayer pairing~\cite{Potter,Pikovski}, and a modified BCS-BEC crossover~\cite{Zinner12,Wang12} are predicted. One dimensional geometries were studied~\cite{Citro08,Kollath, Wang091,Wang092,Dalmonte10}. In Refs.~\cite{Dalmonte11,Wunsch,Zinner} it was shown that the attraction between two dipolar molecules in different one-dimensional systems can overcome the repulsion within each system, leading to stable molecule complexes. In the strong coupling regime, a crystalline structure is predicted~\cite{Citro, Astrakharchik}.

\begin{figure}[t!]
\begin{center}
\includegraphics[width=7.3cm]{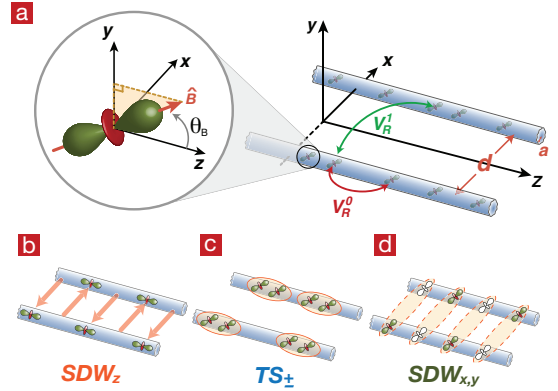}
\caption{(color online) (a) Sketch of an atom with a quadrupole moment, and of two coupled one-dimensional systems. The unit vector $\hat{B}$ indicates the direction of the external magnetic field, along which the quadrupoles are aligned. We assume that $\hat{B}$ is in the $y$-$z$ plane. The angle between $\hat{B}$ and the $z$-axis is $\theta_{B}$. The distance between the two 1D systems is $d$, and $a$ is the length scale of the confinement wave functions. $V^0_R$ and $V^1_R$ denotes the intra- and inter-tube interactions, respectively. In (b), (c) and (d) we sketch the three quantum phases, SDW$_z$, TS$_{\pm}$ and SDW$_{x,y}$, that occur in the phase diagram. SDW$_z$ order corresponds to a spontaneous zig-zag density order, TS$_{\pm}$ corresponds to $p$-wave pairing in each tube. SDW$_{x,y}$ corresponds to an order parameter that consists of one atom in one tube and one hole in the other. These are depicted by a filled and transparent symbol, respectively. We elaborate further on these quantum phases in the text.}
\label{tubes}
\end{center}
\end{figure}

Recently, a new many-body system of ultra-cold atoms was proposed, atomic ensembles interacting predominantly via quadrupolar interactions, Ref.~\cite{Bhongale3}. These can be realized with alkaline-earth atoms, such as Sr, and rare-earth atoms, such as Yb, in the metastable $^3P_2$ states~\cite{Derevianko,Nagel,Santra03,Santra04,Takasu,Jensen,Buchachenko,Stellmer,Sengstock} and in Rydberg-dressed atoms~\cite{Flannery,Tong,Olmos}. A gas of ultra-cold fermionic atoms interacting predominantly via these interactions would thus constitute a quadrupolar Fermi gas. Similar to a dipolar Fermi gases, the higher-order symmetry of the quadrupole-quadrupole interaction results in an exotic competition between pairing and density-wave instabilities in the presence of the Fermi-surface nesting in a two-dimensional optical lattice, as was shown in Ref.~\cite{Bhongale3}. 

In this paper we consider ultra-cold Fermi gases in two coulped one-dimensional systems via quadrupolar interactions, in order to further understand the quantum phases that can be created in quadrupolar gases. While Ref.~\cite{Bhongale3} approached the two-dimensional lattice system with a functional renormalization group calculation, here we address the phase diagram of quadrupolar gases by studying the geometry that is shown in the right panel of Fig.~\ref{tubes}(a). By applying an external magnetic field, the quadrupolar moments of the fermions can be aligned along the angle $\theta_B$ in the $y$-$z$ plane, as illustrated in the left panel of Fig.~\ref{tubes}(a). As a result, the effective intratube and intertube interactions, that emerge from the quadrupole-quadrupole interaction between the fermions, can be controlled by tuning the angle of the external magnetic field. 
 
We determine the quantum phase diagram of the system within a Tomonaga-Luttinger-liquid (TLL)  framework. We employ a pseudo-spin language, in which we formally assign the labels spin-up and spin-down to the two tubes. We bosonize the fermions, as described below, and obtain the low-energy, effective action of the system. From it, we can determine the scaling exponents of the correlation functions of the possible order parameters. By identifying the dominant quasi-long range order, we obtain   the phase diagram as a function of the  distance of two tubes, $d$, and the angle $\theta_B$.
  
We show that there are two magic angles, at which the intratube interactions vanish and change sign. Between the two magic angles, the intratube interaction becomes attractive and the system is described by two gapless TLLs. By further computing various correlation functions, we show that a polarized triplet superfluid (TS$_{\pm}$) and a planar  (pseudo-)spin-density wave (SDW$_{x,y}$), illustrated in Fig.~\ref{tubes}(c) and (d), respectively, compete with each other. Outside of this regime, where intratube and intertube interactions are repulsive, backward scattering between the two tubes create a (pseudo-)spin-gapped state with axial-(pseudo-)spin-density-wave (SDW$_z$) quasi-long range order, sketched in Fig.~\ref{tubes}(b).  

These competing orders are reminiscent of the ones reported in Ref.~\cite{Bhongale3}. There, too, two angles were found at which the interaction between neighboring  sites switched sign. In between these two angles, a bond-ordered solid phase dominated. This order has the same order parameter as the SDW$_{x, y}$ phase that we discuss here: It consists of a particle operator on one site and a hole operator on its neighboring site, orthogonal to the tilting angle. In Ref.~\cite{Bhongale3}, this order parameter developed a checkerboard pattern on a two-dimensional lattice. Due the 1D geometry considered here, it develops a modulation at the wavelength $2 k_{F}$ here, where $k_{F}$ is the Fermi vector of each of the 1D Fermi systems. This order competes with TS$_{\pm}$ pairing, in analogy to the $p$-wave BCS order in Ref.~\cite{Bhongale3}. Outside of the two magic angles, the two-dimensional system develops a checkerboard density-wave order, in analogy to the SDW$_z$ order that we find here. 

However, we note a crucial difference between the 2D half-filled case studied in Ref.~\cite{Bhongale3}, and the two coupled, continuous 1D systems discussed here. For the 2D case, both nesting of the Fermi surface and Umklapp scattering is present.  But while every 1D system generically is nested, a 1D continuous system does not have Umklapp scattering. We indeed find that for a 1D continuous system the pairing instability within each tube is stronger than the $p$-wave pairing tendency in the 2D system at half-filling. We therefore consider an additional commensurate external lattice, which generates Umklapp scattering, and demonstrate that it stabilizes SDW$_{x, y}$ order.

The paper is organized as follows. In Sec. II, we compute the intratube and intertube interactions, the emerge from the bare quadrupolar interaction. Next, we represent the system within Tomonaga-Luttinger-liquid theory in Sec. III. In Sec. IV,  we use a renormalization group calculation to determine the effect of the backscattering term, and calculate the scaling exponents of the correlation functions of the order parameters. Based on these, we determine the  quantum phases diagram. Finally, we present the discussion and summary in Sec. V. 

\section{Quadrupole-Quadrupole interaction in tubes}
Throughout this paper, we assume the fermions to move freely along the direction of the tubes, the $z$-direction, as shown  in  Fig.~\ref{tubes}, and to be confined in the $x$-$y$ plane with transverse wave functions $\Psi_s(\bm{r})=\frac{1}{a\sqrt{\pi}}\exp\left\{\frac{1}{2a^2}\left[(x+sd)^2+y^2\right]\right\}$. $\bm{r}$ is defined as $\bm{r} \equiv (x,y)$. $s=\pm1/2$ is the tube label or the pseudo-spin index, $d$ is the distance between the two tubes, and $a$ is the length scale of the confinement wave functions. The effective 1D Hamiltonian can be represented as 
\begin{eqnarray}\label{H}
\nonumber&&\hspace{-0.8cm}H=\sum_s\int dz\hspace{0.1cm}\psi_{s}^{\dag}(z)\left[\frac{-\hbar^2}{2m}\partial_z^2\right]\psi_s(z)\\
&&\hspace{-0.5cm}+\sum_{s_1,s_2}\int \hspace{-0.1cm}dz_1\int \hspace{-0.1cm}dz_2\hspace{0.1cm}V_R^{|s_1-s_2|}(z_1-z_2)n_{s_1}(z_1)n_{s_2}(z_2),
\end{eqnarray}
where $n_{s}(z)\equiv\psi_s^{\dag}(z)\psi_s(z)$ is the density operator, and $\psi_s(z)$ is the effective 1D single particle operator.  We compute the effective 1D interactions by integrating out the transverse wave functions in the  $x$-$y$ plane as \begin{eqnarray}\label{VR0}
\nonumber&&\hspace{-1cm}V_R^{|s_1-s_2|}(z_1-z_2)=\int d^2\bm{r}_1 \int d^2\bm{r}_2,\\
&&\big|\Psi_{s_1}(\bm{r}_1)\big|^2\left|\Psi_{s_2}(\bm{r}_2)\right|^2V_{R}(\bm{r}_1-\bm{r}_2,z_1-z_2).
\end{eqnarray} 
For $s_1=s_2$, $V^0_R(z)$ is the intratube interaction, and for $s_1\neq s_2$, $V^1_R(z)$ is the intertube interaction. $V_{R}(\bm{R})=V_{R}(\bm{r},z)$ is the quadrupole-quadrupole interaction in real space \cite{Bhongale3}, given by
\begin{eqnarray}\label{QQ}
 V_{R}(\bm{R})=\frac{C_{q}}{R^5}\left[3-30\left(\hat{R}\cdot\hat{B}\right)^2+35\left(\hat{R}\cdot\hat{B}\right)^4\right].
\end{eqnarray}
$\bm{B}$ is an external magnetic field, which can be used to tune the effective interactions $V^0_R(z)$ and $V^1_R(z)$. We denote the strength of the quadrupolar interaction with $C_{q}\equiv3q^2/16$, where $q$ is the quadrupolar moment of an atom or molecule in a 	specific internal state. As an example, we consider the $^3P_2$ state of Yb, with the value  $q\sim30$ [a$^2_0$e] (a$_0$ is the Bohr radius and e is the electronic charge), see Ref. \cite{Buchachenko}, and a corresponding $C_{q}/\hbar$ of  $2\pi\times(4.59\times10^{11})$ [Hz$\cdot$nm$^5$], and  the $^3P_2$ state of Sr, with $q\sim16$ [a$^2_0$e], Ref. \cite{Santra04}, and $C_{q}/\hbar\sim2\pi\times(1.31\times10^{11})$ [Hz$\cdot$nm$^5$]. In the following, we use the value of the quadrupole moment of the $^3P_2$ state of Yb and $a=50$[nm], for specific numerical estimates.

We write the interactions in momentum space, using the Fourier transformation $V_{|s_1-s_2|}(k)=\int_{-\infty}^{\infty}dz\hspace{0.1cm}e^{ikz}V^{|s_1-s_2|}_R(z)$. The intratube interaction is represented as
\begin{eqnarray}\label{intra}
V_0(k)=\frac{C_{q} F_{0}(k) }{3a^4}\left(3-30\cos^2\theta_B+35\cos^4\theta_B\right).
\end{eqnarray}
$F_{0}(k)$ is defined as $F_{0}(k)\equiv\int_{-\infty}^{\infty}F(\tilde{z})e^{ika\tilde{z}}d\tilde{z}$ with
 \begin{eqnarray}
\nonumber&&\hspace{-1cm}F_{0}(\tilde{z})=-\frac{1}{16}\bigg[10\left|\tilde{z}\right|+2\left|\tilde{z}\right|^3\\
&&\hspace{0.5cm}-\sqrt{2\pi}(3+6\tilde{z}^2+\tilde{z}^4)e^{\frac{\tilde{z}^2}{2}}{\rm Erfc}\left(\frac{\tilde{z}}{\sqrt{2}}\right)\bigg],
\end{eqnarray}
where we normalized the $z$ coordinate with the length scale $a$, $\tilde{z}\equiv z/a$.  ${\rm Erfc}(\tilde{z})$ is defined as ${\rm Erfc}(\tilde{z})\equiv 1-{\rm Erf}(\tilde{z})$,  with the error function ${\rm Erf}(\tilde{z})$. We note that the dependence of the bare quadrupolar interaction, Eq.~(\ref{QQ}), and the effective  intra-tube interaction, Eq.~(\ref{intra}), on the angle $\theta_{B}$, is the same, because $\hat{R}\cdot\hat{B} = \cos \theta_{B}$. This is due to the symmetric confinement wave function in the $x$-$y$ plane, where we chose the same length scale $a$ for both the $x$ and the  $y$-direction. 

Except for the case $k=0$, for which we find $F_{0}(k=0)=1$, $F_{0}(k)$ can only be calculated numerically. Similarly, the inter-tube interactions $V_1(k)$ cannot be given in closed form for non-zero momentum ($k\neq0$) and can only be computed numerically. Only at $k=0$, there is a analytical form, given by  
\begin{eqnarray}\label{inter}
\nonumber&&\hspace{-1.cm}V_1(0)=\frac{C_q\sin^4\theta_B}{d^4}\bigg[
\\&&\hspace{0.2cm}4-\frac{1}{12}\bigg(48+24\frac{d^2}{a^2}+6\frac{d^4}{a^4}+\frac{d^6}{a^6}\bigg)e^{-\frac{d^2}{2a^2}}\bigg].
\end{eqnarray} 

\section{Tomonaga-Luttinger-liquid representation}
\begin{figure}[t!]
\begin{center}
\includegraphics[width=6.7cm]{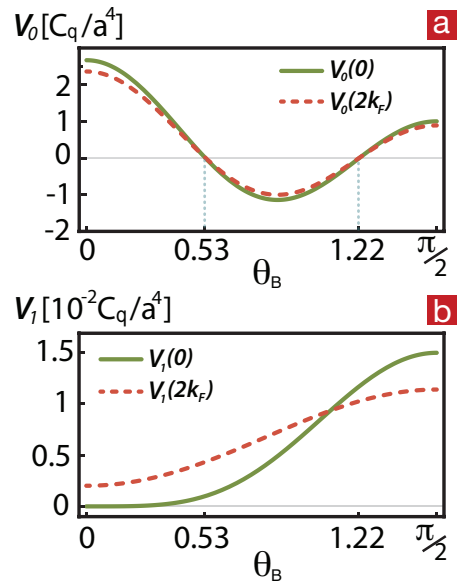}
\caption{(color online) (a) The intra-tube interaction $V_0(k)$ and (b) the inter-tube interaction $V_1(k)$ at momentum $k=0$ and $k=2k_F=n\pi$ versus the angle $\theta_B$ are shown. We choose $d=4a$, the density $n=10^{-3}$[nm$^{-1}$] and $a=50$[nm]. We note  that there are two magic angles $\theta_B^c=0.53$ and $1.22$, where the intra-tube interaction vanishes. }
\label{Vk}
\end{center}
\end{figure}
In this section, we represent the quasi-one dimensional  Hamiltonian in the framework of Tomonaga-Luttinger-liquid theory. We decompose the single-particle operator as $\psi_s(z+z')\sim\sum_{P}\psi_{Ps}(z)e^{iPk_{F}(z+z')}$, where $k_{F}$ is the Fermi vector of each of the tubes, assuming equal density for each of the tubes,  and $P=R/L=+/-$   labels the right-/left- moving fermions, respectively. This representation reduces to the one used for contact interactions, $\psi_s(z)\sim\sum_{P}\psi_{Ps}(z)e^{iPk_{F}z}$, for $z'=0$. Here, however, we deal with non-contact interactions, and the first expression is well suited to identify the correct terms in the TLL action. The kinetic term of the Hamiltonian Eq.~(\ref{H}) is represented as a sum over left- and right-moving fermions with a linear dispersions  $\hbar v_{F} P \left|k - P k_{F}\right|$, see e.g. Ref. \cite{oneD}, where $v_F=\hbar k_F/m$ is the Fermi velocity. The interaction  term of the Hamiltonian is given by 
\begin{eqnarray}\label{HIc}
\nonumber&&\hspace{-0.5cm}H_I=\frac{1}{2}\int dz\sum_{s_1,s_2,P_1,P_2}g_{P_1P_2}^{s_1s_2}\Big[n_{P_1s_1}(z)n_{P_2s_2}(z)\Big]\\
&&\hspace{-0.1cm}+\int dz \sum_{s} V_1(2k_F)\psi_{Rs}^{\dag}(z)\psi_{Ls}(z)\psi_{L\bar{s}}^{\dag}(z)\psi_{R\bar{s}}(z),
\end{eqnarray}
where $\bar{P} \equiv -P$ ($\bar{s} \equiv -s$), $n_{Ps} \equiv \psi_{Ps}^{\dag}\psi_{Ps}$, and the coupling constants are defined as $g_{P_1P_2}^{s_1s_2}=V_{|s_1-s_2|}(0)-\delta_{s_1,s_2}\delta_{\bar{P}_1,P_2}V_0(2k_F)$. The second term of Eq.~(\ref{HIc}) represents back-scattering between particles of different tubes and different chirality~\cite{oneD}. In Fig.~\ref{Vk}(a) and (b), the strength of intra- and inter-tube interaction in units of $C_{q}/a^4$ at  $k=0$ and $k=2k_F$ are plotted versus the angle  $\theta_B$, respectively. We find that there two magic angles $\theta_B^c\simeq0.53$ and $1.22$, where the intra-tube interaction vanishes and changes sign, which corresponds to zeros of $V_{0}(k)$, Eq.~(\ref{intra}). We note that the intra-tube interaction is repulsive for all angles and much smaller than the intra-tube one due to a factor of $1/d^{4}$.

We bosonize the fermions as $\psi_{Ps}(z)=\frac{1}{\sqrt{2\pi\alpha}}e^{i\left[\theta_{s}(z)-P\phi_{s}(z)\right]}$, where $\alpha$ is a short-ranged cutoff and the bosonic fields $\theta(z)$ and $\phi(z)$ are a displacement and a phase field respectively, which are dual to each other. The bosonized representation of the Hamiltonian Eq.~(\ref{H}), which we normalize as  $H_{\rm TL} \equiv H/(2\pi\hbar v_F)$, is  
\begin{eqnarray}\label{TL}
\nonumber &&\hspace{-0.4cm}H_{\rm TL}=\hspace{-0.1cm}\int\hspace{-0.1cm} dz\hspace{-0.1cm}\sum_{j=\rho,\sigma} \hspace{-0.1cm}\frac{v_j}{(2\pi)^2}\left\{K_{j}^{-1}\left[\partial_z\phi_j(z)\right]^2\hspace{-0.05cm}+\hspace{-0.05cm}K_{j}\left[\partial_z\theta_j(z) \right]^2\right\} \\
&&\hspace{3cm}+\frac{g_{1}}{(2\pi)^3\alpha^2}\cos\left[\sqrt{8}\phi_{\sigma}(z)\right],
\end{eqnarray} 
where $\phi_{\rho/\sigma} \equiv \left(\phi_{1}\pm\phi_{2}\right)/\sqrt{2}$, and similarly for $\theta_{\rho/\sigma}$. The backward scattering strength is $g_1=V_1(2k_F)/(2\pi\hbar v_F)$. The Tomonaga-Luttinger parameters are
\begin{eqnarray}
\label{K}K_{\rho/\sigma}&=&\sqrt\frac{1+g_0}{1+2V_{\rho/\sigma}-g_0},\\
v_{\rho/\sigma}&=&\sqrt{\left(1+V_{\rho/\sigma}\right)^2-\left(g_{0}-V_{\rho/\sigma}\right)^2},
\end{eqnarray}
where we denote $V_{\rho/\sigma}\equiv\left[V_0(0)\pm V_1(0)\right]/\left(2\pi\hbar v_F\right)$ and $g_{0}=V_0(2k_F)/(2\pi\hbar v_F)$. 

The resulting Hamiltonian $H_{TL}$ is of the same form as that of a spin-1/2 fermions interacting via contact interactions, see e.g.~\cite{oneD}. Here, however the $SU(2)$ is not present, and therefore the parameters of this Hamiltonian are not subject to this constraint, which allows for additional types of order. Furthermore, the physical interpretation of the quantum phases found here is rather different, because our  pseudo-spin language is merely a notational device. 

The Tomonaga-Luttinger Hamiltonian $H_{TL}$ is a sum  of a Hamiltonian of massless bosons, described by the fields  $\phi_{\rho}$ and $\theta_{\rho}$,  and a sine-Gordon Hamiltonian of the fields $\phi_{\sigma}$ and $\theta_{\sigma}$. 
 We employ a one-loop RG transformation to determine the values of $K_{\sigma}$ and $g_1$ in the low-energy limit. The flow equations of $K_{\sigma}$ and $g_1$ of the sine-Gordon Hamiltonian are 
\begin{eqnarray}\label{RG}
\frac{dK_{\sigma}}{dl}=-2K_{\sigma}^2\left(\frac{g_{1}}{v_{\sigma}}\right)^2,\hspace{0.3cm}\frac{dg_{1}}{dl}=2g_{1}\left(1-K_{\sigma}\right),
\end{eqnarray}
where $l=\ln(\Lambda_0/\Lambda)$ being the logarithm of the ratio between the bare momentum cutoff $\Lambda_0$ and the running cutoff $\Lambda$. These flow equations are perturbative in the backscattering parameter $g_{1}$, which makes this flow most suitable for the weak-coupling limit. For the system we consider in this paper, we indeed find that the interactions are weak in recent experimental setups. For example: for reasonable experimental parameters, such as densities like $n=10^{-3}$, $5\times10^{-4}$ and $2.5\times10^{-4}$[nm$^{-1}$],  for a confinement length scale of $a=50$[nm], and the strength of the quadrupolar interactions $C_{q}$ of the $^3P_2$ state of Yb, mentioned above, we find that the ratio between the quadrupolar interaction and the kinetic energy, $\left(C_{q}/a^5\right)/\left(\hbar^2k_F^2/2m\right)\sim\mathcal{O}(10^{-1})$. This implies that the system is in the weakly interacting limit, which justifies the perturbative RG method used in this study. 
 
Under the RG equations, the bare values of $K_{\sigma}$ and $g_1$ are renormalized to their asymptotic effective values, which we denote as $\tilde{K}_{\sigma}$ and $\tilde{g}_1$, respectively. The backscattering term can either be relevant or irrelevant. If it is irrelevant, the asymptotic values are  $\tilde{g}_1=0$ and $\tilde{K}_{\sigma}>1$ and the Hamiltonian is a sum of  two gapless TLL. If it is relevant,  $g_{1}$ flows to strong coupling, which renormalizes  $\tilde{K}_{\sigma}\rightarrow0$. The resulting effective Hamiltonian has an energy gap in the (pseudo-)spin sector, while the charge sector remains a gapless TLL.

\section{Phase Diagrams}

\begin{figure}[t!]
\begin{center}
\includegraphics[width=8cm]{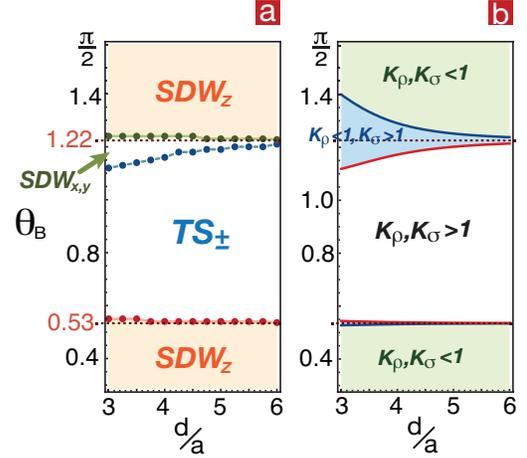}
\caption{(color online) (a) Quantum phase diagram as a function of the distance $d$ of the two tubes and of the angle $\theta_B$. We choose the  density $n=10^{-3}$[nm$^{-1}$] and $a = 50$ nm. The shaded  regime indicates a spin-gapped state.In Fig. (b), we show the corresponding regimes of ($K_{\rho}<1,K_{\sigma}<1$), ($K_{\rho}<1,K_{\sigma}>1$) and ($K_{\rho}>1,K_{\sigma}>1$). }
\label{phase1}
\end{center}
\end{figure}

To determine the quantum phases of the system, we compute the correlation functions of the corresponding order parameters. In the long-distance limit, $z\rightarrow\infty$, these correlation functions behave as $\left\langle O_A(z)O_A(0)\right\rangle\sim z^{-2+\alpha_{A}}$, where $O_A$ is the order parameter and  $\alpha_{A}$ is its scaling exponent. 
 
We consider these order parameters: Change-density wave order is represented by $O_{\rm CDW}=\sum_s\psi_{Rs}^{\dag}\psi_{Ls}$, axial-spin-density wave order by $O_{{\rm SDW}_z}=\sum_{s_1,s_2}\psi_{Rs_1}^{\dag}\sigma^z_{s_1s_2}\psi_{Ls_2}$, and  planar-spin-density wave order by $O_{{\rm SDW}_{x,y}}=\sum_{s_1,s_2}\psi_{Rs_1}^{\dag}\sigma^{x,y}_{s_1s_2}\psi_{Ls_2}$. Furthermore, singlet superfluidity is described by $O_{\rm SS}=\sum_{s}s\psi_{Rs}^{\dag}\psi_{L\bar{s}}^{\dag}$, unpolarized triplet superfluidity by $O_{{\rm TS}_0}=\sum_{s}\psi_{Rs}^{\dag}\psi_{L\bar{s}}^{\dag}$, and polarized triplet superfluidity by $O_{{\rm TS}_{2s}}=\psi_{Rs}^{\dag}\psi_{Ls}^{\dag}$ with $s=\pm1/2$. $\sigma^{i}$, with $i = x,y,z$, are the Pauli matrices. The corresponding scaling exponents are 
  \begin{eqnarray}
  \alpha_{\rm CDW} & = & 2-K_{\rho}-\tilde{K}_{\sigma},\\ 
  \alpha_{{\rm SDW}_{\rm z}} & = & 2-K_{\rho}-\tilde{K}_{\sigma},\\ 
  \alpha_{{\rm SDW}_{\rm x,y}} & = & 2-K_{\rho}-\tilde{K}_{\sigma}^{-1},\\ 
  \alpha_{\rm SS} & = & 2-K_{\rho}^{-1}-\tilde{K}_{\sigma},\\ 
  \alpha_{\rm TS_0} & = & 2-K_{\rho}^{-1}-\tilde{K}_{\sigma},\\ 
  \alpha_{\rm TS_{\pm}} & = & 2-K_{\rho}^{-1}-\tilde{K}_{\sigma}^{-1}
  \end{eqnarray}
respectively, see e.g.~\cite{oneD}. The quantum phases are determined by the most slowly decaying correlation function, i.e. the largest $\alpha$. Here, as it is typical for 1D systems at zero temperature, only  quasi-long-ranged order is achieved, rather than true long ranged order. In the spin-gapped state, for which $\tilde{K}_{\sigma}\rightarrow0$, only two phases are possible, SDW$_z$ and SS. Thus, the quantum phase is determined by comparing the corresponding scaling exponents $2-K_{\rho}$ and $2-K_{\rho}^{-1}$,  respectively~\cite{footnote}. In the regime, in which backscattering is irrelevant, a more intricate competition of orders appears, as we see below.

\begin{figure}[t!]
\begin{center}
\includegraphics[width=8cm]{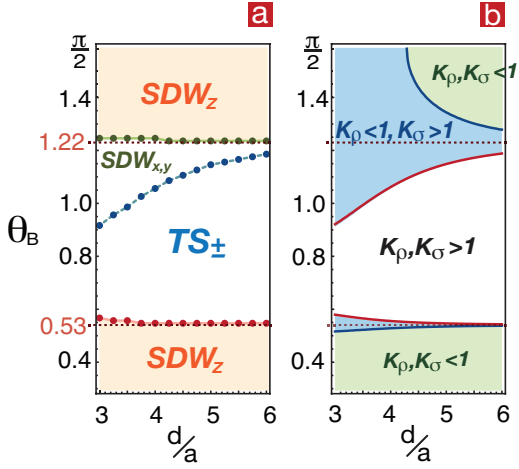}
\caption{(color online) (a) Quantum phase diagram and (b) the corresponding regimes of ($K_{\rho}<1,K_{\sigma}<1$), ($K_{\rho}<1,K_{\sigma}>1$) and ($K_{\rho}>1,K_{\sigma}>1$) for a lower density than Fig. \ref{phase1}, specifically   $n=5\times10^{-4}$[nm$^{-1}$]. We again choose $a = 50$ nm.}
\label{phase2}
\end{center}
\end{figure}

\begin{figure}[t!]
\begin{center}
\includegraphics[width=8cm]{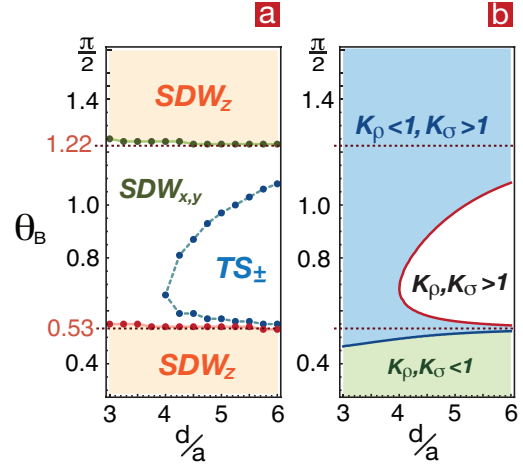}
\caption{(color online) (a) Quantum phase diagram and (b) the corresponding regimes of ($K_{\rho}<1,K_{\sigma}<1$), ($K_{\rho}<1,K_{\sigma}>1$) and ($K_{\rho}>1,K_{\sigma}>1$) for density $n=2.5\times10^{-4}$[nm$^{-1}$], and $a = 50$ nm.}
\label{phase3}
\end{center}
\end{figure}

We now map out the full quantum phase diagram numerically, as a function of the distance of two tubes, $d$, and and the angle $\theta_B$ in Fig.~\ref{phase1}(a), \ref{phase2}(a) and \ref{phase3}(a) for the atomic densities $n=10^{-3}$, $5\times10^{-4}$ and $2.5\times10^{-4}$ [nm$^{-1}$] respectively.
We choose $a=50$[nm] and $C_{q}=2\pi\hbar\times(4.59\times10^{11})$ [$\hbar$Hz$\cdot$nm$^5$]. We find that for attractive intratube interaction, between the two magic angles, the backscattering term is irrelevant, and the spin sector remains gapless, while for repulsive  intratube interactions  a spin-gapped state is generated. This  transition almost coincides with the magic angles of the intratube interactions. 
 
We calculate the phase boundary analytically from the RG equations: By integrating Eq.~(\ref{RG}) and assuming  $\tilde{K}_{\sigma}=1$ and $\tilde{g}_1=0$, we find that the phase boundaries in terms of the bare values are $1-1/K_{\sigma}-\ln(K_{\sigma})=-(g_{1}/\alpha_{\sigma})^2/2$. Since the system is in the weak coupling regime, we expand this expression, using Eqs.~(\ref{K}), which reduces this condition to $V_0(0)-V_0(2k_F)=V_1(0)-V_1(2k_F)$. However, the intertube interactions are all positive and much smaller than intratube ones, as illustrated in Fig.~\ref{Vk}. Therefore, to leading order, the phase boundaries of the spin-gapped state are $V_0(0)-V_0(2k_F)\simeq0$, which  occurs near the magic angles. Because the two intertube interaction is weak in all three cases depicted, the phase boundaries do not vary significantly in Figs.~\ref{phase1}(a), \ref{phase2}(a), \ref{phase3}(a). 

In the spin-gapped regime, i.e. $\theta_B<0.53$ or $\theta_B>1.22$, the dominant quantum phase only depends on the bare value of $K_{\rho}$. We illustrate the regimes of ($K_{\rho}<1,K_{\sigma}<1$), ($K_{\rho}<1,K_{\sigma}>1$) and ($K_{\rho}>1,K_{\sigma}>1$) in Fig.~\ref{phase1}(b), \ref{phase2}(b) and \ref{phase3}(b), accompanying the corresponding phase diagrams.  $K_{\rho}<1$ implies SDW$_z$ is favored over SS. The order parameter of SDW$_z$ is $O_{{\rm SDW}_z}=\psi_{Rs}^{\dag}\psi_{Ls}-\psi_{R\bar{s}}^{\dag}\psi_{L\bar{s}}$, implying an antiferromagnetic pattern of the pseudospin. This corresponds to a  zigzag patten of the atomic  densities of the two tubes, as shown in Fig.~\ref{tubes}(b),

In the regime of irrelevant backscattering, TS$_{\pm}$ and SDW$_{x,y}$ compete with each other. Based on their scaling exponents, the boundary between these two phase is at $K_{\rho}=1$, or $V_0(2k_F)-V_0(0)=V_1(0)$ from Eq.~(\ref{K}). Thus, TS$_{\pm}$ and SDW$_{x,y}$ emerges for $K_{\rho}>1$ and  $K_{\rho}<1$, respectively. This leads to the phase diagrams shown in Fig.~\ref{phase1}, \ref{phase2} and \ref{phase3}. We also find  that with decreasing density, and thus increasing interaction strength, the regime of SDW$_{x,y}$ grows. In Fig.~\ref{tubes}(c), we sketch the quantum phase of the polarized triplet superfluid, where particles (or holes) pair up in the same tube due to the intra-tube attraction. For the quantum phase SDW$_{x,y}$, described by  $O_{{\rm SDW}_{x}}=\psi_{Rs}^{\dag}\psi_{L\bar{s}}+\psi_{R\bar{s}}^{\dag}\psi_{Ls}$, correlations between a particle and a hole on different tubes is sketched  in Fig.~\ref{tubes}(d). The intertube correlations of SDW$_{x,y}$ resulting in an interference pattern in momentum space has recently been proposed as a measurement in a time-of-flight experiment~\cite{Wang091}. 

\section{Discussion}
Comparing to quadrupolar Fermi gases in a two-dimensional lattice, as discussed in Ref. \cite{Bhongale3}, we find similar features of the phase diagram. Outside of the two magic angles, the two-dimensional system develops a checkerboard density-wave order, in analogy to the SDW$_z$ order that we find here. Between the two magic angles, a bond-ordered solid phase and $p$-wave BCS order compete with each other, in analogy to the competition of SDW$_{x,y}$ and TS$_{\pm}$ that we find here. 

However, a qualitative difference between the two cases is that in the two-dimensional lattice at half-filling the fermions can undergo Umklapp scattering. Since we consider a continuous one-dimensional model in the present study, density wave instabilities are weaker due to the absence of a commensurate lattice. Indeed we note that in two dimensions the bond ordered solid phase always dominates if the quadrupoles are tilted along one of the lattice directions, see Ref. \cite{Bhongale3}, and that $p$-wave BCS order  appears only if they are tilted way from the lattice directions. 
    
We therefore  expect that SDW$_{x,y}$  can be further stabilized by applying an external potential with period $4k_F$ along the two tubes, i.e. $V_{\rm ext}\propto\cos\left(4k_Fz\right)$ to induce Umklapp processes~\cite{oneD,Robinson, zWang}. The Umklapp term in bosonized form is
\begin{eqnarray}
H^u_{\rm TL}=\frac{g_{u}}{(2\pi)^3\alpha^2}\cos\left[\sqrt{8}\phi_{\rho}(z)\right],
\end{eqnarray}
with $g_{u}$ being the strength of the Umklapp coupling~\cite{oneD,Hu}.
 This gives a sine-Gordon Hamiltonian for the fields $\phi_{\rho}$ and $\theta_{\rho}$. The RG equations for $K_{\rho}$ and $g_{u}$ are the same as the equations for $K_{\sigma}$ and $g_{1}$~\cite{oneD,Hu}, in particular 
\begin{eqnarray}\label{RGu}
\frac{dK_{\rho}}{dl}=-2K_{\rho}^2\left(\frac{g_{u}}{v_{\rho}}\right)^2,\hspace{0.3cm}\frac{dg_{u}}{dl}=2g_{u}\left(1-K_{\rho}\right).
\end{eqnarray}
The bare values of $K_{\rho}$ and $g_u$ will be renormalized under these flow equations to their asymptotic effective values.

\begin{figure}[t!]
\begin{center}
\includegraphics[width=8cm]{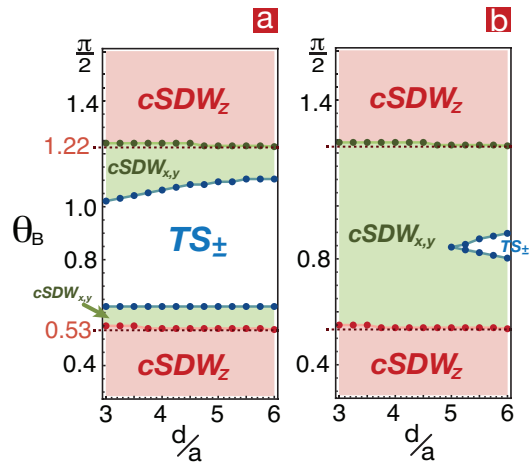}
\caption{(color online) 
 Quantum phase diagram for $n=10^{-3}$[nm$^{-1}$]  and $a = 50$ nm, in the presence of  Umklapp scattering.
 For the strength of  Umklapp scattering we choose $g_u=0.02$ in  (a) and  $g_u=0.05$ in (b).}
\label{phase4}
\end{center}
\end{figure}

In Figs.~\ref{phase4}(a) and (b) we show the resulting phase diagram, for density $n=10^{-3} $[nm$^{-1}$] and for the  two bare values $g_u=0.02$ and $g_u=0.05$, respectively. In comparison to the phase diagram without Umklapp interaction, see Fig.~\ref{phase1}(a), we find that the commensurate lattice enhances density-wave instabilities and suppresses pairing. Moreover, we find that in the regime of the density instability, the flow of $g_{u}$ to the strong coupling regime renormalizes $K_{\rho}$ to zero. This implies a charge-gapped phase, Ref.~\cite{oneD}. We therefore have now a SDW$_{z}$ and a SDW$_{x,y}$ phase for which $K_{\rho}$ is zero, in contrast to the previous case without the periodic potential for which $K_{\rho} >0$. To distinguish these, we denote these phases  as commensurate spin-density waves (cSDW) with a charge gap, i.e. cSDW$_{z}$ and cSDW$_{x,y}$. We note that the phase boundary between cSDW$_{z}$ and cSDW$_{x,y}$ is not modified, because it is determined by $\tilde{K}_{\sigma}$ only. The boundary between cSDW$_{x,y}$ and TS$_{\pm}$ however is modified by the commensurate lattice, as shown in  Figs.~\ref{phase4}(a) and (b). The SDW$_{x,y}$ regime grows for increasing $g_{u}$, until it entirely dominates over  TS$_{\pm}$.

A natural system to compare quadrupolar gases to, are dipolar gases. The quantum phases of coupled 1D dipolar systems have been discussed in Ref.~\cite{Wang091}. The phase diagrams of dipolar and quadrupolar fermions exhibit a similar structure. For repulsive intratube interactions  a spin-gapped state arises. For attractive intratube interactions, SDW$_{x,y}$ competes with TS$_{\pm}$. Compared to dipolar fermions, quadrupolar fermions display an additional spin-gapped regime, due to the higher-order symmetry of the quadrupole-quadrupole interaction. Moreover, since intertube interactions of quadrapolar gases are significantly smaller than dipolar gases, the phase boundaries of the spin-gapped states essentially coincide with the magic angles. 

In conclusion,  we have investigated the  quantum phase diagram of quadrupolar Fermi gases in two coupled one-dimensional systems. Within the framework of Tomonaga-Luttinger-liquid theory, and using the one-loop RG transformation in the weak-coupling limit, we have determined the phase diagram as a function of the distance between the two tubes and the alignment angle of the quadrupolar moments. We show that the phase transitions to a spin-gapped state coincide with the two magic angles of the intratube interaction, at which the interaction vanishes and changes signs. In the spin-gapped state,  SDW$_z$ quasi-long-range order dominates, which corresponds to a zigzag pattern of the atomic densities of the two tubes. Outside of the spin-gapped regime,  we show that SDW$_{x,y}$ and TS$_{\pm}$ order compete with each other. TS$_{\pm}$ order corresponds to pairing within each tube, whereas SDW$_{x,y}$  corresponds to particle-hole pairing between the two systems. We demonstrate that this intriguing order can be further enhanced by applying an external periodic potential that is commensurate with the density of the atoms, to induce the Umklapp scattering. In particular, this would be the order that dominates for a half-filled lattice system, in analogy to the bond order solid phase that was found in Ref.~\cite{Bhongale3}.

\begin{acknowledgments}
We thank Y.-P. Huang and H.-H. Lin for useful discussions.
 We acknowledge support from the Deutsche Forschungsgemeinschaft through
the SFB 925 and the Hamburg Centre for Ultrafast Imaging, and from the Landesexzellenzinitiative Hamburg, which is supported by the Joachim Herz Stiftung.

\end{acknowledgments}



\begin{thebibliography}{999}
\bibitem{footnote}
Although the exponents of CDW and SDW$_z$ are the same, SDW$_z$ is favored due to the logarithmic corrections of the amplitude of the correlation function, see Ref.~\cite{oneD}.
\end{thebibliography}
\end{document}